\documentclass[prd,preprint,superscriptaddress,floatfix,eqsecnum,nofootinbib,12pt]{revtex4}

\usepackage{amssymb,amsmath,amsthm,graphicx,ulem}
\pdfoutput=1
\usepackage{graphicx,subfigure}
\usepackage{ulem}
\usepackage{epsfig}
\usepackage{amsmath}
\usepackage{amsfonts}
\usepackage{amssymb}
\usepackage[usenames]{color}
\usepackage[letterpaper,left=2.cm,right=2.cm,top=2.5cm,bottom=2.5cm]{geometry}
\usepackage[T1]{fontenc}

\newcommand{\beq}{\begin{equation}}
\newcommand{\eeq}{\end{equation}}
\newcommand{\be}{\begin{equation}}
\newcommand{\ee}{\end{equation}}
\newcommand{\beqa}{\begin{eqnarray}}
\newcommand{\eeqa}{\end{eqnarray}}
\newcommand{\beqar}{\begin{eqnarray*}}
\newcommand{\eeqar}{\end{eqnarray*}}
\newcommand{\bea}{\begin{eqnarray}}
\newcommand{\eea}{\end{eqnarray}}

\def\qf{}

\newcommand{\ra}{\rightarrow}

\newcommand{\nn}\nonumber
\newcommand{\eqn}[1]{(\ref{#1})}
\begin{document}

\title{Further Holographic Investigations of Big Bang Singularities}

\author{Netta Engelhardt}
\email{engeln@physics.ucsb.edu}
\affiliation{Department of Physics, UCSB, Santa Barbara, CA 93106, USA}
\author{Thomas Hertog}
\email{thomas.hertog@fys.kuleuven.be}
\affiliation{Institute for Theoretical Physics, KU Leuven, 3001 Leuven, Belgium}
\author{Gary T. Horowitz}
\email{gary@physics.ucsb.edu}
\affiliation{Department of Physics, UCSB, Santa Barbara, CA 93106, USA}

\vskip 1in
%\date{\today}

\bibliographystyle{unsrt}

\begin{abstract}
\noindent 
We further explore the quantum dynamics near {\qf past} cosmological singularities in anisotropic Kasner-AdS solutions using gauge/gravity duality. The dual description of the bulk evolution involves ${\cal N}=4$ super Yang-Mills on the contracting branch of an anisotropic de Sitter space and is well defined. We compute two-point correlators of Yang-Mills operators of large dimensions using spacelike geodesics anchored on the boundary. The correlator between two points separated in a direction with negative Kasner exponent $p$ always exhibits a pole at horizon scales, in any dimension, which we interpret as a dual signature of the classical bulk singularity. This indicates that the geodesic approximation selects a non-normalizable Yang-Mills state.

\end{abstract}

\maketitle 
 
%%%%%%%%%%%%%%%%%%%%%%%%%%%%%%%%%%%%%%%%%%%%%%%%%%%%%%%%%%%%%%%%%%%%%%%%%%

%\baselineskip16pt

%\tableofcontents

\section{Introduction}

A truly predictive theory of cosmology requires an understanding of the past singularity, in order to explain how the distinctive features of our universe emerged from the early quantum gravitational phase and why they are what they are. A central issue one would like to understand in this context is how a classical spacetime arises from the singularity. Since our usual notions of space and time are likely to break down near cosmological singularities, a natural approach to study this problem is to find a dual description of the early universe in terms of more fundamental variables. 

In string theory we do not yet have a dual description of realistic cosmologies. The AdS/CFT correspondence, however, allows us to describe and study singularities in toy model cosmologies that are asymptotically anti-de Sitter (AdS), in terms of a dual quantum field theory living on the conformal boundary. The first examples of such `AdS cosmologies' were constructed in \cite{Hertog:2004rz,Hertog:2004b}. These were solutions of $N=8$, $D=4$ supergravity involving only gravity and a single scalar field where smooth, spherically symmetric asymptotically AdS initial data evolves into (and from) a singularity which extends all the way out to infinity. Models of this type were further explored in \cite{Turok:2007ry,Barbon:2011ta,Smolkin:2012er} and other models were studied e.g. in \cite{Craps:2006xq,Das:2006dz,Awad:2008jf}.

For the AdS cosmologies in \cite{Hertog:2004rz}, {\qf it was shown that if one defines the dual on the global AdS boundary, the field theory also becomes singular when the bulk singularity hits the boundary} \cite{Hertog:2004b,Turok:2007ry}. However, if one views the dual field theory as living on a de Sitter (dS) boundary then it remains everywhere well defined \cite{Maldacena:2010un,Harlow:2010my}. This is because even though the bulk scalar field turns on a (homogeneous) negative mass deformation in the dual, the conformal coupling to the dS boundary geometry ensures the deformed theory is stable, at least for sufficiently small deformations. 

From a dual dS viewpoint the bulk singularity lies in the infinite future (or past) and therefore corresponds to an asymptotic field theory state. While this is appealing in some sense, it also raises the question whether the dual on dS captures the physics associated with the singularity. This point was sharpened recently when it was found that the probes which are best understood, such as extremal surfaces which end on the boundary, stay well away from the high curvature region near the singularity \cite{Engelhardt:2013tra}. 

To identify more explicitly dual signatures of cosmological singularities, we have recently constructed a new class of five-dimensional  AdS cosmologies in which the bulk is a vacuum, anisotropic Kasner-AdS space that emerges from an initial singularity \cite{Engelhardt:2014mea}. The dual description of the bulk evolution  is simply ${\cal N}=4$ super Yang-Mills on the contracting branch of deformed (anisotropic) de Sitter space and is again well defined. In contrast with the isotropic solutions discussed above, here there do exist bulk geodesics with endpoints on the boundary which come close to the singularity. Specifically, for boundary separations in a direction with a negative Kasner exponent $p$, the corresponding bulk geodesics bend {\it towards} the singularity in the interior.

In the large $N$ limit of the dual field theory, the leading contribution to the two-point correlator of an operator $\mathcal{O}$ of high conformal dimension $\Delta$ is often approximated  by the (regulated) length of spacelike bulk geodesics anchored on the boundary. 
%\footnote{When there is no Euclidean continuation defining the state, the geodesic approximation is not guaranteed to give an accurate representation of the two-point correlator under consideration. In this paper, we assume it does give an accurate approximation.}.
 Hence the existence of such geodesics probing the high curvature region near the singularity in Kasner-AdS opens up the possibility of using the dual conformal field theory to study the quantum dynamics near singularities\footnote{See e.g. \cite{Fidkowski:2003nf,Festuccia:2005pi} for attempts to probe the singularity inside AdS black holes using geodesics with endpoints on the boundary.}. In \cite{Engelhardt:2014mea} we computed the equal time two-point correlator in the geodesic approximation for the particular case of points separated in a $p=-1/4$ direction. We found that it indeed has distinct features which, we argued, encode information about the bulk expanding cosmology: at horizon separation the correlator has a pole, and at large distances it decays as a power law with a power that depends on the local expansion rate. 

In this paper we extend the analysis in \cite{Engelhardt:2014mea} in several ways. We show that the pole at the horizon scale is a direct  result of the singularity, and occurs generically in all cases when $ p<0$. This pole in the correlator is a result of bulk geodesics which probe the region of large curvature near the singularity and get pulled out to the boundary.  As the boundary separation approaches the horizon size, the geodesics approach a null geodesic that lies entirely on the boundary. Approximately null geodesics naturally lead to a pole because their length diverges much more slowly compared to the length of spacelike geodesics. After regularization, this yields a pole in the two-point function. 

We also discuss the pole from the standpoint of the dual field theory. It turns out that a two-point correlator with standard short-distance bethavior --- as modeled by ours --- can only diverge at spacelike separation if the state is not normalizable. To our knowledge, this is the first example where the geodesic approximation fails to reproduce the two-point function in any normal state  in the dual field theory. We argue that this is a direct result of the bulk singularity, and discuss some possible interpretations of this state.

%The pole shows up as a divergence in the momentum space correlator. We interpret this as a manifestation in the dual that the classical gravity approximation in the bulk that we used to compute the two-point function breaks down near curvature singularities. On the other hand we find that the structure of the momentum space correlator indicates the pole should be absent at weak coupling in the dual field theory. We therefore expect that the pole will be smoothed out when stringy corrections are taken into account. However its physical meaning, if any, in a more realistic cosmology remains to be understood.

In addition to the cosmological singularity, our model has two more subtle singularities which we will discuss.  We show that neither singularity affects our results about the pole, and that there are slightly more sophisticated AdS cosmologies in which they are absent altogether. 

An outline of this paper is as follows. We start in the next section by reviewing the bulk solutions of interest. We then proceed in section III to solve for the geodesics that determine the two-point correlator. In section IV we show that there is a pole at the horizon scale for all $p<0$ resulting from geodesics that get close to the singularity. To illustrate these general arguments, we  work out a simple example of a 5+1 dimensional bulk solution whose Kasner exponents are $\pm 1/2$ in section V. The following section contains our discussion of the pole from the standpoint of the dual field theory. In section VII, we discuss the two more subtle singularities in our model and how to remove them. This is followed by some concluding remarks  in section VIII.

\section{Set-up \label{setup}}

\noindent Consider the following bulk metric \cite{Das:2006dz}
\begin{equation} \label{KasnerAdS}
 ds^{2}= \frac{1}{z^{2}}\left ( -dt^{2} + \sum\limits_{i} t^{2p_{i}}dx_i^{2}  +dz^2 \right)
 \end{equation}
 where we have set the AdS radius to 1.
 When $\sum\limits_{i} p_{i} = 1 = \sum\limits_{i} p_{i}^{2}$, this is a solution to Einstein's equation with negative cosmological constant. In less than five {\qf spacetime} dimensions, the only possible values for $p_i$ are $0,1$ and the metric is equivalent to pure AdS. In \cite{Engelhardt:2014mea} we studied the five dimensional case. This will be our main focus here, although  we will also consider some higher dimensional examples. When there is an exponent $p_i$ which is neither vanishing nor 1, there is a curvature singularity at $t=0$. {\qf In this case there is always one negative Kasner exponent in five dimensions, and at least one in higher dimensions.} 
 
 The natural dual to the five-dimensional solution is super Yang-Mills on the Kasner metric: $ds^2 = -dt^{2} + \sum\limits_{i} t^{2p_{i}}dx_i^{2} $. Since our goal is to try to learn about the bulk singularity, it is more {\qf appealing} to work in a conformal frame in which the boundary metric is nonsingular. This can be achieved by pulling out a factor of $t^2$ in the above metric so the new conformal factor is $t/z$. Setting $t = e^\tau$, the resulting metric on the
 boundary is now an anisotropic version of de Sitter space
\be\label{anidS}
ds^2 =  -d\tau^2 + \sum_i e^{-2(1-p_i) \tau} dx_i^2,
\ee
where $(1-p_{i})$ may be viewed as the Hubble parameter in the $x_{i}$ direction. In addition to the obvious translational symmetries, \eqn{KasnerAdS} is invariant under a dilation symmetry:
\be 
z \ra  \lambda z, \quad t \ra  \lambda t,  \quad x_i \ra \lambda^{(1-p_i)} x_i
\label{dilation}
\ee
This leaves the new conformal factor $t/z$ invariant and thus acts as an isometry of the boundary metric \eqn{anidS}.

The simplest way to probe the singularity is by studying the two-point correlator of a high conformal dimension $\Delta$ operator in the dual $SU(N)$ gauge theory.
In the semiclassical bulk (large $N$) limit, the leading order contribution to this two-point correlator is often {\qf specified} by the length of spacelike bulk geodesics connecting the two points:
\begin{equation} \left \langle\psi \right | \mathcal{O}\left(x\right)\mathcal{O}\left(x'\right)\left | \psi \right\rangle =e^{-\mathcal{L}(x,x') m}\end{equation}
\noindent where $\left | \psi \right\rangle$ is the state of the CFT, $m$ is the mass of the bulk field corresponding to the boundary operator $\mathcal{O}$, and $\mathcal{L}(x,x')$ is the {\qf (regularized)} length of the bulk geodesic. When $\mathcal{O}$ is a scalar operator, $m$ and $\Delta$ are related via the following relation:
\begin{equation} \Delta = \frac{d}{2} + \sqrt{\frac{d^{2}}{4} + m^{2}}\end{equation}
\noindent where $d$ is the {\qf boundary} spacetime dimension, so in the limit of large conformal dimension, $\Delta \approx m$.

The length of these geodesics is naturally infinite, since they travel to the boundary at infinity. This divergence is normally regulated by truncating the geodesics at some  cutoff  $z = \epsilon$, which corresponds to implementing a UV cutoff at energy scale $\frac{1}{\epsilon}$ in the dual field theory. The $\epsilon$-independent contribution to the correlator is then extracted by subtracting the divergent contribution of geodesics in pure AdS from the length. This standard regularization scheme must be modified due to our  nonstandard conformal factor. We want the cutoff to correspond to a fixed proper length on the boundary. When this proper length is small, the corresponding bulk radial cutoff can be found by looking at the domain of dependence (in the boundary metric) of a ball with that diameter. One then finds the bulk point such that an outgoing radial null geodesic reaches the tip of this domain of dependence (both to the future and past).  This is satisfied with a cutoff at constant $\tilde z = z/t$. This can be checked directly, but it follows from the fact that near each point on the boundary, the spacetime looks like AdS, and a constant UV cutoff in pure AdS corresponds to a constant value of the conformal factor.

Note that the only component of the holographic computation of the correlator that depends on the choice of conformal factor is the regularization scheme. Bulk geodesics may therefore be computed without reference to the coordinate $\tilde{z}$, at least prior to integration of the length functional. We will follow this technique and solve for geodesics in the $z$ coordinate system, and regulate the lengths with respect to $\tilde{z}$. 

\section{Geodesics in Kasner-AdS$_{d+1}$}

We are interested in equal time correlators, so we consider geodesics anchored on some boundary time slice, $t=t_{0}$, and let their endpoints be separated in only one spatial direction. We shall  take this direction to be $x_{1}$, hereafter referred to as $x$; we shall similarly henceforth refer to $p_{1}$ as $p$. Translation symmetry in $x_{2}$ and $x_{3}$ allows us to fix these to constants along the geodesic, so the geodesic travels in a $(2+1)$-dimensional space with effective metric given by:
\begin{equation} \label{metric3d}
ds^{2} = \frac{1}{z^{2}}\left(-dt^{2} + dz^{2} +t^{2p}dx^{2} \right).\end{equation}
\noindent The geodesic endpoints at $z=0$ are $\{t=t_{0}, \ x=\pm x_{0}\}$, and the proper boundary separation (in the de Sitter conformal frame) is therefore $\mathcal{L}_{bdy}= 2x_{0}t_{0}^{p-1}$. Changing the boundary condition $t=t_{0}$ to $t=t_{1}=\lambda t_{0}$ is equivalent, by the dilation symmetry of the conformal boundary  \eqref{dilation}, to changing $x=\pm x_{0}$ to $x=x_{1}=\pm\lambda^{p-1}x_{0}$. This transformation leaves the proper boundary separation invariant: $\mathcal{L}_{bdy}(t_{1},x_{0})=\mathcal{L}_{bdy}(t_{0}, x_{1})$. The  time slice value $t_{0}$ may therefore be fixed to some convenient value, say $t_{0}=1$, and $\mathcal{L}_{bdy}$ changed by varying the value of $x_{0}$. 

We find the geodesics by extremizing the length functional:
\begin{equation} L = \int\sqrt{\frac{t^{2p} dx^{2}- dt^{2} +dz^{2}}{z^{2}}}. \label{functional}\end{equation}
Note that $L$, and by extension the geodesics found by extremizing $L$, are independent of the spacetime dimension. While calculations below are executed using time as the parameter along the geodesic, it is instructive to first examine the equations for the geodesics using the spatial direction $x$ as the parameter. The equation for the time propagation $t(x)$ of the geodesic decouples from the equation for $z(x)$, and is given by:

\begin{equation} p\frac{t(x)^{2 p} - 2 t'(x)^{2}}{ t(x)} + t''(x)=0. \label{paramx} \end{equation}

Since the endpoints are at equal time, there must be a turning point where $t'(x_\star) = 0$. Near this point,  \eqref{paramx} is  approximately
\begin{equation}t(x) t''(x) = -p t(x)^{2p}.\label{turningpt}
\end{equation}
\noindent Restricting to positive time (so the $t=0$ singularity is in the past), we find that negative values of $p$ require $t''(x_\star) > 0$, while positive values of $p$ require $t''(x_\star) < 0$. Geodesics that propagate in a direction with a negative Kasner exponent must therefore be attracted to the singularity, while those with positive Kasner exponent must be repelled from it.\footnote{It might appear that when $p = 1/2$, \eqref{turningpt} can be satisfied by a geodesic that turns around at the singularity, so $t(x_\star) = 0$ and $t''(x_\star)>0$. But one can show that this geodesic does not reach the boundary with finite $\mathcal{L}_{bdy}$.}

While it is perhaps more intuitive to parametrize the geodesic in terms of $x$ and solve for $t(x)$ using \eqref{paramx}, parametrizing the geodesics in terms of $t$ rather than $x$ significantly facilitates the computation. We therefore first solve for $x(t)$, by using an inversion technique on the equations for $t(x)$  \eqref{paramx}, and then subsequently use $x(t)$ to solve for $z(t)$. The equations of motion, obtained by extremizing the length functional \eqref{functional} and parametrizing the geodesics in terms of $t$ are:
\begin{align} & x''(t)t=px'(t) \left [ t^{2p}x'(t)^{2}-2\right] \notag\\
&z''(t) z(t) =1-z'(t)^{2}-t^{2p-1} x'(t)^{2}\left [ t- p z(t)z'(t)\right]\label{eom}
\end{align}

By defining a new variable $ u(t)= t'(x)^{2}$, we may easily solve eq. \eqref{paramx} and find 
\be\label{solnu}
u(t) = t^{2p} + c t^{4p},
\ee
 where $c$ is an integration constant. Recall that $t'(x)$ vanishes, or equivalently $u(t)$ vanishes at the turning point. This implies that for $t>0$, the turning point occurs at $t_{*}= (-c)^{-1/2p}$. The function $x(t)$ can be obtained from $u(t)$ by direct integration; the result below agrees with the earlier work of 
 \cite{Banerjee:2015fua}\footnote{Ref. \cite{Banerjee:2015fua} solved Eqs. \eqref{eom} for a different set of boundary conditions (and with fewer simplifications).}:

\begin{align} \label{xsoln}& x(t) = \int \frac{1}{\sqrt{u(t)}}dt\\
& = \frac{ (-c)^{\frac{1}{2}-\frac{1}{2 p}} \sqrt{\pi } \Gamma\left[\frac{1}{2} \left(3+\frac{1}{p}\right)\right]}{\left(1-p^2\right) \Gamma\left[1+\frac{1}{2 p}\right]}+\frac{t \sqrt{t^{2 p}+c t^{4 p}} \left(-(1+p) t^{-2 p}+c\ \phantom{}_{2}F_{1}\left[1,1+\frac{1}{2 p},\frac{1}{2} \left(3+\frac{1}{p}\right),-c t^{2 p}\right]\right)}{p^{2}-1}.\notag \end{align} 

Here we have used the fact that the endpoints are by construction symmetric about the reflection $x\rightarrow - x$ to fix the integration constant. The above expression must be accompanied by an important caveat: when $p= - \frac{1}{2n+1}$, where $n$ is an integer, the hypergeometric function above  is not well defined\footnote{This subtlety was not discussed in \cite{Banerjee:2015fua}.}. We will defer until the end of the section to solve the equations for these values of the Kasner exponent.

It follows from \eqref{eom} and \eqref{solnu} that the equation for $z(t)$ can be written
\begin{equation}
 z''(t) z(t) + z'(t)^{2} + \frac{t-p z(t)z'(t)}{t+ c t^{1+2p}}=1.
\end{equation}
\noindent To solve for $z(t)$, we define a new variable $v(t)=z'(t)z(t)$. Then the equation simplifies: 
\begin{equation} 
\frac{t-p v(t)}{t+c t^{1+2 p}}+v'(t)=1
\end{equation} 
\noindent This first order differential equation can be integrated, yielding:
\begin{equation} 
v(t) = \frac{c_{3} t^p}{\sqrt{1+c t^{2 p}}}+\frac{c t^{1+2 p} \phantom{}_{2}F_{1}\left[1,1+\frac{1}{2 p},\frac{1}{2} \left(3+\frac{1}{p}\right),-c t^{2 p}\right]}{1+p}
\end{equation}
\noindent where $c_{3}$ is an integration constant. Geodesics that contribute to the correlator must be smooth. This is a nontrivial constraint at the turning point: we must require that $dz/dx = z'(t)/x'(t)=0$ as $t$ approaches the turning point at $t=(-c)^{-1/2p}$, or equivalently $v(t)/x'(t)=0$. This fixes $c_{3}$ to the following value:
\begin{equation} c_{3} = -\frac{2 \left(-\frac{1}{c}\right)^{\frac{1}{2}+\frac{1}{2 p}} c \sqrt{\pi } \Gamma\left[\frac{1}{2} \left(3+\frac{1}{p}\right)\right]}{(1+p)\Gamma\left[1+\frac{1}{2 p}\right]}\end{equation}
Finally, the bulk function $z^{2}(t)$ can now be obtained by integrating $v(t)$:
\begin{align} 
z^{2}(t) & = \frac{2 \left(-c\right)^{\frac{p-1}{2 p}} c \sqrt{\pi } t^{1+p} \Gamma\left[\frac{1}{2} \left(3+\frac{1}{p}\right)\right] \phantom{}_{2}F_{1}\left[\frac{1}{2},\frac{1+p}{2 p},\frac{1}{2} \left(3+\frac{1}{p}\right),-c t^{2 p}\right]}{(1+p)^2 \Gamma\left[1+\frac{1}{2 p}\right]}\notag \\
& +\frac{c t^{2+2 p} (1+ct^{2p})\phantom{}_{2}F_{1}\left[1,1+\frac{1}{2p},\frac{1}{2} \left(3+\frac{1}{p}\right),-c t^{2 p}\right]^2}{(1+p)^2}+ c_{4} \label{zsoln}
\end{align}
\noindent where we have made use of various identities of hypergeometric functions to simplify the expression. Here $c_{4}$ is an integration constant determined by requiring that geodesics reach the boundary at $t_{0}=1$: 

\begin{align} 
c_{4} = &\frac{c}{(1+p)^2} \Bigg (\frac{4 \left(-\frac{1}{c}\right)^{\frac{1+p}{2 p}} p\sqrt{\pi } \Gamma\left[\frac{1}{2} \left(3+\frac{1}{p}\right)\right] \phantom{}_{2}F_{1}\left[\frac{1}{2},\frac{1+p}{2 p},\frac{1}{2} \left(3+\frac{1}{p}\right),-c\right]}{\Gamma\left[\frac{1}{2 p}\right]}\notag \\
& -(1+c)\phantom{}_{2}F_{1}\left[1,1+\frac{1}{2 p},\frac{1}{2} \left(3+\frac{1}{p}\right),-c\right]^2\Bigg ) 
\end{align}
The entire solution is now expressed in terms of one free parameter, $c$, which determines the  boundary separation of the endpoints.  From \eqref{solnu} it is clear that as $c \to -1$, the turning point in the geodesic where $t'(x) = 0$ approaches $t=1$. This is the same value of $t$ as the endpoints. Since \eqref{paramx} shows that  the geodesic cannot stay at constant $t$, this is  consistent only if the distance between the endpoints vanishes in this limit.

Note that when the Kasner exponent takes the form $p=-1/2n$ for integer $n$, the useful identity 
\begin{equation}  _{2} F_{1} \left (1, 1-n, \frac{3}{2}-n; z\right) = \sum\limits_{j=0}^{n-1} (-1)^{j} \begin{pmatrix} n-1 \\j  \end{pmatrix} \frac{\Gamma(\frac{5}{2}-n-j)}{\Gamma (\frac{5}{2}-n)}z^{j}\end{equation}
simplifies the expressions for both $x(t)$ and $z^{2}(t)$ into finite-order polynomials in $- c t^{2p}$. We shall provide an explicit example in Section \ref{OneHalf} for $p=-1/2$. The original example of $p=-1/4$ may be found in \cite{Engelhardt:2014mea}. 

We conclude this section by treating the separate case in which $p = - \frac{1}{2n+1}$. By setting $p = - \frac{1}{2n+1}$ in the equations of motion \eqref{eom}, we may implement a variable redefinition $y= - t^{-2p}c^{-1}$, which allows us to obtain a solution for these particular values of $p$ via the same methodology described above:
\begin{align}
 x(t) & =  \frac{t^{1-p} \sqrt{1+ct^{2p}}}{1-p} \left ( \phantom{}_{2}F_{1} \left [ 1, \frac{1}{2}- \frac{1}{2 p}; 1- \frac{1}{2p}; - t^{-2p}c^{-1}\right]-1 \right) + \frac{i (-c)^{\frac{1}{2} - \frac{1}{2p}}\sqrt{\pi} \Gamma\left[1- \frac{1}{2p} \right] }{(1-p) \Gamma\left[\frac{1}{2}-\frac{1}{2p} \right]}\\
z^{2}(t)& = t^2\phantom{}_{2} F_{1}\left[\frac{1}{2},-\frac{1}{2 p},1-\frac{1}{2 p},-\frac{t^{-2 p}}{c}\right]^{2}+\frac{c_{3} t \sqrt{c+t^{-2 p}} \phantom{}_{2} F_{1}\left[1,\frac{-1+p}{2 p},1-\frac{1}{2 p},-\frac{t^{-2 p}}{c}\right]}{c} + c_{4},
\end{align}
\noindent where the constants $c_{3}$ and $c_{4}$ are determined as above:
\begin{align} c_{3}& = \frac{i (-c)^{\frac{p-1}{2 p}}  \sqrt{\pi }\Gamma \left[1-\frac{1}{2 p}\right]}{\Gamma\left[\frac{p-1}{2 p}\right]}\\
c_{4}& = -\frac{1}{2} \phantom{}_{2}F_{1} \left[\frac{1}{2},-\frac{1}{2 p},1-\frac{1}{2 p},-\frac{1}{c}\right]^{2}+\frac{i (-c)^{-\frac{1+p}{2 p}} \sqrt{1+c} \sqrt{\pi } \Gamma \left[1-\frac{1}{2 p}\right] \phantom{}_{2}F_{1}\left[1,\frac{-1+p}{2 p},1-\frac{1}{2 p},-\frac{1}{c}\right]}{\Gamma\left[\frac{-1+p}{2 p}\right]}
\end{align}

It is not possible, given the complicated form of the solutions, to compute the regulated length of these geodesics for general  values of the exponent $p$. We can, however, extract some properties of the two-point correlator from the behavior of the solutions alone. 

\section{Pole at Horizon Size Separations for General $p<0$}

As discussed in the introduction, a primary motivation for our investigation is to translate the problem of curvature singularities from the language of gravity to that of quantum field theory. To do so in this setup, it is imperative to identify a definitive property of the two-point correlator that is a clear signature of the bulk initial singularity. In \cite{Engelhardt:2014mea}, we found that for boundary separation in a direction with Kasner exponent $p=-1/4$, the corresponding two-point correlator features a pole at a separation that is precisely equal to the cosmological horizon. In this section, we will show that this pole arises generically {\qf for all} $p < 0$. As mentioned earlier if the surface at $t=0$ is a curvature singularity {\qf then there is always at least one $p<0$ direction.}

Before we proceed to the calculation, we first explain why this pole is a direct signature of the bulk singularity. As the boundary separation approaches the horizon scale from above, there are spacelike bulk geodesics which get closer and closer to the cosmological singularity. These geodesics also approach the boundary, so the limiting curve is a null geodesic lying entirely on the boundary and ``bouncing'' off past infinity of the anisotropic de Sitter space.  Spacelike geodesics shot in from infinity do not usually stay close to the boundary. The reason they do so in our case is a direct result of the singularity. One can view the singularity as ``dragging the tip of the geodesic out to infinity''. 

As explained earlier, the regulated length of each geodesic is computed by introducing a UV cutoff at small $\tilde z = z/t$ and subtracting the usual divergence of a geodesic in pure AdS (which stays on a constant time slice). Since our bulk geodesics are becoming null, their length diverges much more slowly than a standard bulk geodesic; after subtracting the usual divergence, then, the regulated length is large and negative. This produces a pole in the two-point function. For a fixed cutoff, the entire bulk geodesic eventually lies past the cutoff, so the pole is capped off at some boundary separation slightly larger than the horizon. However, as the cutoff goes to zero, the pole is recovered.

Below we use the geodesic solutions for general $p$ of the previous section to argue that the two-point correlator separated in a direction with negative Kasner exponent $p$, \textit{for any negative value of $p$} and any spacetime dimension, will feature a pole at the cosmological horizon. Thus whenever there is a genuine curvature singularity, there is {\qf at least} one direction along which the correlator diverges at the horizon. To do this, we show that spacelike bulk geodesics always approach a null boundary geodesic for $p<0$. Note that \cite{Fidkowski:2003nf} studied a case in which the bulk spacelike geodesics approached a null {\it bulk} geodesic and argued that it did not contribute to the correlator. We will argue below that in our case, they do contribute to the correlator and the pole is physical.

From the general form of $x(t)$ and $z(t)$ derived above (\ref{xsoln},\ref{zsoln}), it is clear that at $c=0$, the following is  a solution to the geodesic equation for any value of $p$:\footnote{This is also a solution when $p = -1/(2n+1)$.}
\begin{align} &x(t) = \frac{t^{1-p}}{1-p} \\ 
& z(t)=0. \label{nullgeod}\end{align}
\noindent The boundary separation of the geodesic endpoints is then $2/(1-p)$, which is precisely the size of the cosmological horizon in the $x$ direction. This geodesic is a two-part null geodesic which bounces off past infinity at $t=0$. Below we show that while this solution is an isolated geodesic for $p>0$, it is a limit of a sequence of spacelike bulk geodesics for $p<0$. 

One of the boundary conditions imposed on these geodesics is that the geodesics be smooth at the turning point, or equivalently $1/x'(t)=0$ {\qf at $t=t_{*}$}. From \eqref{solnu} we see that when the turning point does not occur at the singularity, this is achieved at 
\be
 t_\star^{-2p} = - {c}
 \ee
\noindent  Since the bulk spacetime only includes $t>0$, $t_\star$ must be positive and therefore $c$ must be negative. For positive Kasner exponent, as $c$ approaches zero (from below) the turning point of the geodesics approaches $t=\infty$, rather than at the big bang singularity, so the null geodesic in Eq. \eqref{nullgeod} must be an isolated solution. This is in agreement with the fact that spacelike bulk geodesics with $p>0$ curve away from the singularity, in contrast with geodesics with $p<0$, which curve towards it. When the Kasner exponent is negative, the turning point of spacelike bulk geodesics approaches the singularity at $t=0$. It is therefore possible that the null boundary geodesic at $c=0$ is precisely the limit of a set of spacelike bulk geodesics, as we found in \cite{Engelhardt:2014mea} for $p=-1/4$. 

In order to determine definitively whether the $c=0$ solution exists as a limit of a continuous set of spacelike bulk geodesics for $p<0$,  consider the series expansion of $x(t)$ and $z(t)$ around $c\sim 0$. 
\begin{align} & z(t) = \frac{\sqrt{1-t^{2(1+p)}}}{1+p} (-c)^{1/2} +\mathcal{O}((-c)^{3/2})\\
& x(t) = \frac{t^{1-p}}{1-p} - \frac{t^{1+p}}{2(1+p)}c+\mathcal{O}(c^2)
\end{align}
\noindent  From the series expansion of $z^{2}(t)$, it is thus clear that $z$ approaches 0 as $c$ approaches 0, so bulk geodesics with $p<0$ approach a boundary geodesic as we take $c$ to zero. The zeroth order contribution to this solution is therefore precisely the null boundary geodesic at $c=0$. The first order term approaches 0 as $c$ approaches zero, and subsequent orders again approach 0, so we conclude that spacelike bulk geodesics with $c<0$ approach a null boundary geodesic as $c$ approaches zero. 

We now address the question of whether the geodesics that give rise to the pole in fact ``see'' regions of large curvature, as would be necessary if we are to claim that the pole is a singularity-related phenomenon. Since the pole in the two-point correlator is a result of geodesics that approach the singularity as they approach the boundary, it may a priori not be clear that the geodesics approach the singularity sufficiently rapidly to probe any region of high curvature. To ascertain whether the curvature diverges as the geodesics approach the null boundary geodesic, we  compute the curvature of spacetime at the turning point of the geodesic as the turning point approaches the singularity at $t=0$, or equivalently as $c$ approaches zero and the spacelike bulk geodesics approach the null boundary geodesic.  The Kretschmann scalar in five dimensional Kasner-AdS is given in terms of the coordinates in \eqref{KasnerAdS} by
\begin{equation} \label{kret}R_{abcd}R^{abcd} = 40 -16(p-1)p^{2} \frac{z^{4}}{t^{4}}.\end{equation}
where $p$ denotes any of the $p_i$'s.\footnote{When the three $p_i$ satisfy the Kasner conditions, $\sum\limits_{i} p_{i} = 1 = \sum\limits_{i} p_{i}^{2}$, the expression $(p_i -1)p_i^2$ is independent of $i$.}
The value of the Kretschmann scalar at the turning point of a geodesic is therefore a function of $z(t_{*})^{4}/t_{*}^{4}$. The leading order contribution to $z(t_{*})^{4}/t_{*}^{4}$ near $c=0$ for negative Kasner exponent is $(-c)^{2(1+1/p)}$, which diverges as $c$ approaches zero. So the Kretschmann scalar diverges as $t_{*}^{-4(1+p)}$ when the turning point of the geodesic approaches the singularity. As the geodesics approach the boundary, then, they do so in a way that  probes regions of progressively larger curvature.

Finally, we turn to the question of which geodesics contribute to the correlator. 
Since the two-point function on the boundary is the limit of a  two-point function in the bulk, the choice of geodesics depends on a choice of bulk state for linearized quantum fields on our background. 
Recall that the geodesics are labelled by the parameter $c$, and $c \to -1$ is the coincidence limit while $c\to 0$ (from below) gives the pole at the horizon.
The geodesic approximation is never justified when the spacetime is not analytic. In the cases we have studied,  geodesics with $c > 0$ go through the singularity, and they  can also give rise to unphysical divergences (see \textit{e.g.} \cite{Engelhardt:2014mea}). So we will not include these geodesics.
Among the geodesics we keep (with $c<0$), there can be more than one with the same boundary separation ${\cal L}_{bdy} $. In such cases, we sum over all contributions.\footnote{Unlike the setup of~\cite{Fidkowski:2003nf}, our geometry does not have an analytic continuation to a Euclidean metric. Thus we have no natural way to specify the quantum state. Related to this, it is difficult to justify more rigorously which geodesics contribute in a steepest descent approximation.
}

   A previous attempt to  study  the singularity inside an eternal black hole \cite{Fidkowski:2003nf} found a sequence of spacelike geodesics which approached a null bulk geodesic which touched the singularity. However it was argued that those geodesics did not contribute to the correlator since they yielded unphysical results. This does not apply to our case: the null limiting geodesic in question lives on the boundary, not in the bulk.The following argument strongly suggests that these geodesics must be included in any geodesic approximation to the correlator.
 In all cases we have studied, when the boundary separation ${\cal L}_{bdy} $ is slightly larger than the horizon, there are two spacelike geodesics  for the same ${\cal L}_{bdy} $. As ${\cal L}_{bdy} $ approaches the horizon scale, one geodesic approaches a null boundary geodesic, while the other remains spacelike. It may at first sight seem that the inlcusion of only the latter geodesic could be a simple way of eliminating the pole altogether; however, as ${\cal L}_{bdy} $ increases, these two geodesic families merge and then become complex; an example will be provided in the next section. Since the two-point function is real at spacelike separations, complex geodesics can only contribute jointly with their complex conjguates. Including only one of the real geodesics {\qf for values of ${\cal L}_{bdy} $ below the merger point} would result in a discontinuity in the correlator at some  length scale larger than the horizon. This unphysical result  strongly suggests that both real geodesics contribute and the pole at the horizon is physical. 
 Finally we note that the resulting pole in the correlator occurs along a null surface (the horizon) rather than a spacelike surface and therefore is not ruled out by causality.

\section{The $p_{1}=-1/2$ Case: \label{OneHalf}}

\begin{figure}[h]
\begin{center}
\includegraphics[width=3in]{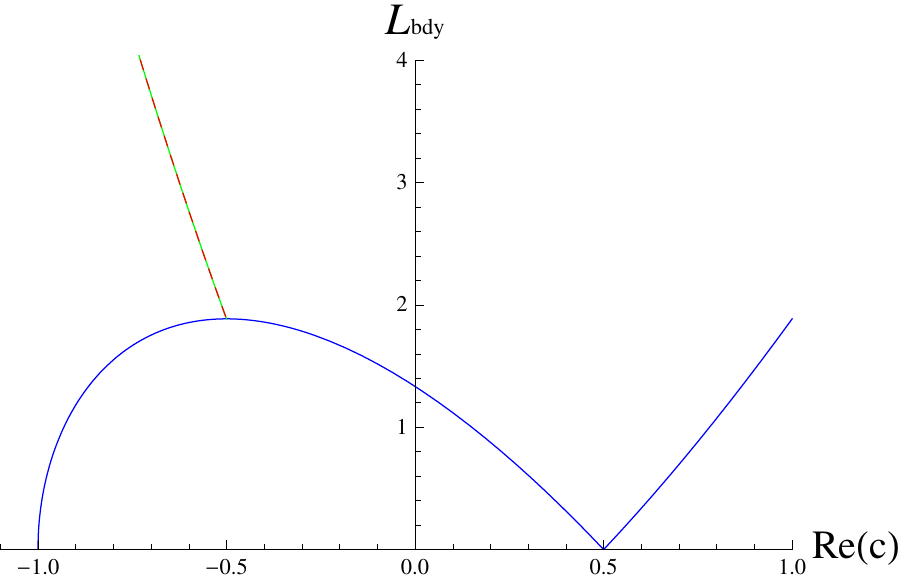}
\caption{The boundary separation in terms of the real part of the parameter $c$.  Solutions with real $c$ are in solid blue and those with complex $c$ are in dashed red and green. Since geodesics with $c>0$ go through the singularity where the spacetime is non-analytic, we do not include them in the geodesic approximation of the two-point correlator.}
\label{bdysep}
\end{center}
\end{figure}

To illustrate the above analysis in an explicit setup, we consider a particularly simple example with a 5+1-dimensional bulk, and Kasner exponents $p_{1}=-1/2$, $p_{2}=p_{3}=p_{4}=1/2$. The equations for the geodesics are unchanged by the number of dimensions, so the same solutions obtained for general $p$  apply to these particular values. The hypergeometric functions simplify considerably in this case and the solution is:
\begin{align}
x(t) & = \frac{2}{3} (-2 c+t) \sqrt{c+t}  \label{xfor1/2}\\
z(t) & = 2 \sqrt{c (t-1)} 
\end{align}
\noindent where we have set the boundary endpoints at $t=1$, and the turning point is at $t_\star =-c$. The parameter $c$ controls the boundary separation which is given by ${\cal L}_{bdy} = 2x(1) = (4/3)(1-2c)\sqrt{1+c}$. There are three values of $c$ corresponding to a given ${\cal L}_{bdy} $; one of these values is always real, and the remaining two are either real or complex conjugates. These three solutions are illustrated in Fig. \ref{bdysep}. 

When $c$ is real and positive, $z$ is complex. But the singularity occurs at $t=0$ for all complex $z$ \eqref{kret}, and the turning point $t_{*}=-c$ occurs at negative time, so such geodesics go through the singularity.  As discussed above, we do not include the $c\geq 0$ geodesics in the two-point correlator. We do, by contrast, include all geodesics with Re$(c)<0$. It may be \textit{a priori} unclear why both complex conjugates with Re$(c)<0$ must be included. This follows from the fact that a two-point correlator at spacelike separation must be real. The geodesics with complex conjugate $c$'s yield complex conjugate contributions to the two-point correlator. Since the correlator must be real, we must sum over both complex conjugates. As discussed in the previous section, this also provides an indication that both real geodesics with $c<0$ must be included.

%\textbf{Presumably we want to take this out: Because Kasner-AdS is non-analytic at the big bang singularity, the geodesic approximation to the two-point correlator breaks down at $t=0$ \cite{Louko:2000tp}. We must therefore exclude any geodesics which go through the singularity from our computation of the correlator, and will therefore only consider contributions to the two-point correlator from  Re $c \le 0$.}   

As discussed in Section \ref{setup}, boundary-anchored geodesics in asymptotically (locally) AdS spacetimes  have infinite length. For this reason, it is standard to implement a short-distance cutoff by computing the length of the geodesics up to some small value $\epsilon$ of $\tilde{z}=z/t$, and then subtracting the logarithmic divergent contribution from pure AdS.
Since geodesics in our setup are parametrized by $t$, the regulator is implemented by computing the length of the geodesic up to a fixed value of $t=1-\delta^{2}$ rather than $t=1$. This is then converted to the corresponding value of $\tilde{z} = z(t)/t$. For $p=-1/2$, the $\tilde{z}$-cutoff $\epsilon$ is related to the $t$-cutoff $\delta^{2}$ by $\delta = \frac{\epsilon}{2 \sqrt{-c}}$. Note that in our case, the UV cutoff excludes geodesics that come very close to the null boundary geodesic causing the pole. The lengths of these geodesics is entirely within the regime removed by the UV cutoff, and they therefore do not contribute to the regulated correlator. 
 
The length of the geodesic is given by:
\begin{align*} L & = 2\int_{-c}^{1-\delta^2} \sqrt{\frac{x'(t)^2/t - 1 +z'(t)^2}{z(t)^2}} dt\\
&= 2  \int\limits^{1-\delta^{2}}_{-c} \sqrt{\frac{1+c}{c+t}}\frac{dt}{1-t} \\
& = 2 \left .  \text{Arctanh}\left[\sqrt{\frac{c+t}{1+c}}\right]\right |^{1-\delta^{2}}_{-c} \\
& = \ln 2 -\ln\left[-\frac{1}{8 c (1+c)}\right] +2 \ln \epsilon +\mathcal{O}(\epsilon^{2})\end{align*}
\noindent Subtracting the divergent pure AdS contribution $2\ln \epsilon$ and neglecting $\mathcal{O}(\epsilon)^{2}$ contributions finally yields the two-point correlator along the $p=-1/2$ direction
\begin{equation}\left \langle \mathcal{O}(-x)\mathcal{O}(x)\right\rangle = \left(-\frac{1}{16 c (1+c)}\right)^{\Delta}\label{p=-1/2}
\end{equation}
\noindent This correlator has precisely two points of divergence: $c=-1$, corresponding to the usual short-distance singularity $\sim 1/{\cal L}_{bdy}^{2\Delta}$ and $c=0$, corresponding to the null boundary geodesic at horizon separation $\sim  1/({\cal L}_{bdy}- \mathcal{L}_{\text{hor}})^{\Delta}$. 

When ${\cal L}_{bdy}$ becomes large, \eqref{xfor1/2} implies that ${\cal L}_{bdy} \propto c^{3/2}$, so the above correlator falls off as ${\cal L}_{bdy}^{-4\Delta/3}$. This is consistent with the  conjecture made in \cite{Engelhardt:2014mea} that for general $p$, the large distance fall-off is ${\cal L}_{bdy}^{-2\Delta/1-p}$. This holds in all cases we can check, but we do not yet have a general derivation.

\section{Correlators in the Dual Field Theory}

We now ask what can be said about the state in the dual CFT that gives rise to the singularity at the horizon in the two-point function.  This state must respect the translation and dilation symmetry of the background since our correlator does. It must also have the standard short distance singularity.
It turns out that standard quantum field theory in curved spacetime does not allow the two-point function to diverge at spacelike separation in any normal state.\footnote{We thank S. Hollands and D. Marolf  for discussions on this point, and S. Hollands for providing the argument that follows.} More precisely, if operators commute at spacelike separation, and the two-point function has the usual short distance behavior and is positive in the sense that $\langle \mathcal{O}(f)\mathcal{O}(f)\rangle \ge 0$ where $f$ is any smearing function, then the two-point function cannot blow up at finite separation. The rough argument is the following. Let $\chi$ be any real test function localized near a point $x=0$ in local coordinates. Then the usual short distance behavior implies that
\be
\langle \mathcal{O}(\chi e^{ikx})\mathcal{O}(\chi e^{ik'x})\rangle \to 0
\ee
as $k,k' \to \infty$ unless $k= -k'$ and $k'$ is a future directed timelike vector. This is just the statement that the usual short distance singularity is positive frequency. If $f_k$ and $g_{k'}$ are two test functions of the above type localized around two  separate points, then the Schwarz inequality implies
\be
\langle{\cal O}(f_k){\cal O}(g_{k'})\rangle \le \langle{\cal O}(f_k){\cal O}(f_k)\rangle^{1/2}  \langle{\cal O}(g_{k'}){\cal O}(g_{k'})\rangle^{1/2}
\ee
The right hand side vanishes in the limit of large $k,k'$ unless $k'$ is a 
future-directed timelike vector and $k$ is a past-directed timelike vector. If the points are spacelike separated, the operators commute, so the left hand side is equal to $\langle{\cal O}(g_{k'}){\cal O}(f_k)\rangle$. This vanishes whenever $k$ is not future-directed. The net result is that the two-point function always vanishes when $k, k'$ are large.  If the Fourier transform of a function vanishes for all large $k$, the function is not singular. 

The conclusion is that our state must fail to satisfy positivity $\langle \mathcal{O}(f)\mathcal{O}(f)\rangle \ge 0$, and hence is not normalizable in the usual sense. 
This appears to be the first  example where the geodesic approximation fails to reproduce the correlator in any normalizable state. It is natural to trace this failure to the existence of the singularity in the bulk. As shown above, the bulk geodesics responsible for the pole probe the high curvature region of spacetime close to the singularity and get pulled out toward the boundary. By contrast in directions where the Kasner exponent is positive, the geodesics bend away from the singularity and the correlator is perfectly smooth except for the usual short distance divergence.

\begin{figure}[t]
\begin{center}
\includegraphics[width=2.3in]{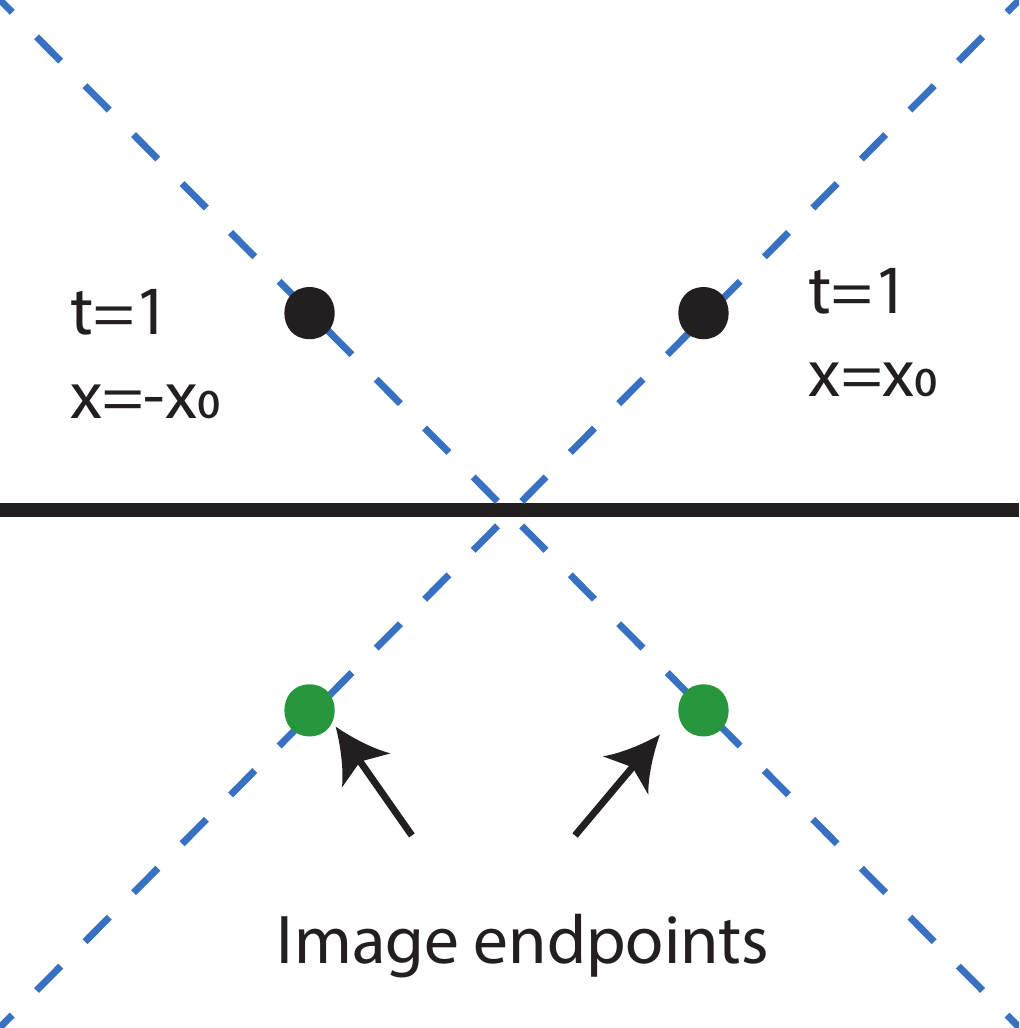}
\caption{A boundary conformal theory setup, in which one imposes Dirichlet boundary conditions at $t=0$ and computes the two-point correlator at $\pm x_{0}$ and $t=1$ via the method of images.}
\label{BCFT}
\end{center}
\end{figure}

There is a simple intuitive picture of the state described by the geodesic approximation. It is reminiscient of correlated massless quasi-particles are produced at each point in space at past infinity. As quasi-particles propagate away from each other, they are always separated by the horizon scale.
An intuitive picture of how {\qf a state of this kind} might be defined {\qf in the dual} can be obtained as follows\footnote{We thank Tom Hartman for pointing this out.}:
In the Kasner frame, the surface at $t=0$ is a boundary to spacetime on which the field theory lives. Because the theory is $\mathcal{N}=4$ super Yang-Mills, the setup is a 3+1-dimensional boundary conformal field theory (BCFT) in a curved upper-half space, where the initial data at the $t=0$ surface determines the state of the field theory at all times. Unfortunately, the number of symmetries in curved spacetime is insufficient to fix or even significantly narrow down the structure of the two-point correlator. Despite the differences between conformal field theory in 1+1 dimensions and in higher dimensions, it is illuminating to consider a 1+1 dimensional boundary conformal field theory. Consider, then, a scalar field in a 1+1 dimensional CFT on a half-plane with a boundary at $t=0$ and Dirichlet boundary conditions at $t=0$ on the scalar field theory, see e.g. Fig. \ref{BCFT}. 

In this case, the method of images may be used to compute the equal-time two-point correlator of the scalar field \cite{DiFrancesco}. When the separation between the two points is smaller than the distance between one of the points and the boundary at $t=0$, the two-point correlator is not sensitive to the effects of the boundary. When the separation between the points is precisely the same as the separation from one point to the boundary --- i.e. at the horizon --- the points are null-separated from the mirror images. The correlator therefore features a lightcone singularity precisely at horizon separation. Subsequently the correlator decays at large separations. This setup and the resulting physics are quite similar to the phenomena we have observed in 3+1-dimensional BCFT, {\qf at strong coupling,} when one considers $t=0$ as the initial boundary of spacetime. 
%{\qf On the other hand the arguments above suggest that the weakly-coupled field theory may not have a pole. If so this would mean that the state changes at finite coupling\footnote{{\qf This is easier to imagine by considering the time reversed bulk solutions, which exhibit a big crunch singularity in the future. It is then intuitively clear that evolution at weak coupling may well produce a very different state dual to the singularity}.}.}

%More speculatively, one can imagine that {\qf our choice of} bulk state at the big bang singularity can {\qf also be implemented} in terms of Dirichlet boundary conditions. Then the bulk two-point correlator may feature a pole at horizon separations as well. This is because on constant $z$ slices, the geodesic equation is solved by a null geodesic that bounces off the singularity; if the singularity implies Dirichlet boundary conditions, we may expect that the bulk two-point function can also be computed via the method of images\footnote{The bulk theory is admittedly not a CFT.}, {\qf which would lead} to a pole when the separation points lie on the lightcone\footnote{Such a state would not be Hadamard.}. 
%{\qf Vice versa, the absence of a pole in} the weakly-coupled field theory {\qf would then mean} that the initial conditions of a quantum, stringy singularity cannot be modeled as local Dirichlet boundary conditions.

\section{Subtle Singularities}

In addition to the cosmological singularity at $t=0$, our model has two more subtle singularities. In this section we show why these do not affect our results and, in fact, can both be removed.

\subsection{Poincare horizon}

Once we put a nonflat metric on the constant radial slices in the bulk,  there is a singularity at the Poincare horizon, $z = \infty$. This is  to the future of the $t=0$ singularity and can  be viewed as a ``big crunch" singularity. There is also a simple extension of our model which removes this singularity completely. One simply adds an extra compact direction, and starts with the six-dimensional AdS soliton metric \cite{Witten:1998zw,HorowitzMyers}:
\begin{equation} 
 ds_{\text{soliton}}^{2} = \frac{1}{z^{2}} \left [ \left(1-\frac{z^{5}}{z_0^5}\right) d\theta^{2} + \eta_{\mu\nu} dx^{\mu}dx^{\nu} +  \left(1-\frac{z^{5}}{z_0^5}\right)^{-1} dz^{2} \right] \label{SolitonMetric}
\end{equation}
This metric can be obtained from the standard planar black hole by a double analytic continuation. If $\theta$ is periodic with period $4\pi z_0/5$, the circle smoothly caps off at $z = z_0$.   One can again replace the flat metric with any Ricci flat metric and still satisfy Einstein's equation with a negative cosmological constant. Using the Kasner metric we get \cite{Engelhardt:2013jda}
\begin{equation}
 ds_{\text{KAS}}^{2}= \frac{1}{z^{2}}\left [\left(1-\frac{z^{5}}{z_0^5}\right) d\theta^2 -dt^{2} + t^{2p_{1}}dx_1^{2} + t^{2p_{2}} dx_{2}^{2} + t^{2p_{3}}dx_{3}^{2}  +\left(1-\frac{z^{5}}{z_0^5}\right)^{-1} dz^{2}  \right]\label{Kasneton}
  \end{equation}
This metric has a cosmological singularity at $t=0$, just like our earlier model, but no longer has a Poincare horizon.

We now replace the conformal factor $1/z^2$ with $H^2 t^2 /z^2$ where $H$ is a constant, and divide the metric in brackets by $H^2 t^2$. Writing $Ht = e^{H\tau}$, $x_i = H^{p_i}y_i$, our boundary metric  looks like an anisotropic version of five-dimensional de Sitter:
\be\label{anidSS}
ds^2 =  -d\tau^2 + \sum_i e^{-2(1-p_i) H\tau} dy_i^2 + e^{-2 H\tau} d\theta^2
\ee
The horizon scale is ${\cal L}_{hor} = 2[(1-p_i)H]^{-1}$. One can now repeat the calculation of the equal time correlator. Since the bulk geodesics effectively live in a three-dimensional space, the only effect of the extra $\theta$ direction is the modification to $g_{zz}$. Since the pole at the horizon scale comes from geodesics that stay close to the boundary ($z=0$), this result is completely unchanged. The large distance behavior will eventually be affected, since these geodesics probe deep into the bulk and will eventually notice that the circle is capped off. This happens when the endpoints are separated at a scale that can be called the confinement scale. This confinement scale clearly becomes infinite as $z_0 \rightarrow \infty$. So one can choose the free parameters $z_0$ and $H$ so that there is a wide range of distances which are larger than the horizon scale but smaller than the confinement scale. In this range, our earlier result about the large distance fall-off of the correlator will still hold.

\subsection{de Sitter horizon}

 We now return to our previous five dimensional bulk solution. Let us order the exponents so $p_1$ is the smallest, and write the metric on the boundary 
 \be\label{bdydS}
 ds^2 =  -d\tau^2 + \sum_i e^{-2H_i\tau} dy_i^2 
 \ee
 with $H_i = (1-p_i)H$.
When the expansion rates are all equal, the surface $\tau = \infty$ is a smooth null surface and the spacetime can be extended to an expanding phase in the future. However, when they are different, the spacetime cannot be extended. Curvature invariants do not blow up as $\tau\rightarrow \infty$ since the dilation symmetry ensures that they are all time independent, but tidal forces do blow up showing that there is a (null) curvature singularity at  $\tau = \infty$.

To see this, consider two nearby null geodesics in the $(\tau,y_1)$ plane with tangent vector $\ell$ which are separated in the $y_2$ direction.   By translational symmetry, $K = \ell \cdot \partial/\partial y_1$ is constant, so 
\be
\dot \tau = K e^{H_1 \tau}
\ee
where a dot denotes derivative with respect to an affine parameter $\lambda$. This is easily integrated to yield
\be
e^{-H_1\tau} = 1- H_1 K \lambda
\ee
By translational symmetry the geodesics stay at constant $y_2$, so their proper separation is 
\be
D(\lambda) = D_0 e^{-H_2\tau} = D_0 (1-H_1 K \lambda)^{H_2/H_1}
\ee
where $D_0$ is the initial separation. Since the smallest exponent $p_1$ cannot be positive and the other exponents cannot be negative, $H_2/H_1 = (1-p_2) / (1-p_1) \le 1$. It follows that the relative acceleration between the two geodesics, $\ddot D$, diverges as $\tau \rightarrow \infty$ (unless $p_1 =0$ which is the nonsingular Milne case). This indicates infinite tidal forces and a curvature singularity.

One can remove this singularity by choosing a late time, $\tau_f$, and letting our exponents $p_i$ become time dependent after this time in such a way that they are all equal by $\tau = 2\tau_f$. Then the boundary metric will be exactly de Sitter after this time and can be extended into the expanding phase. The bulk solution will only be changed to the causal future of $z = 0, \tau = \tau_f$, and none of the geodesic calculations discussed earlier will be affected.

\section{Discussion}

Our ultimate goal is to study quantum gravitational effects near cosmological singularities using holography. To do this, we have found an example of a cosmological singularity with a well defined holographic dual. This is the Kasner-AdS bulk solution which (in five bulk dimensions) is dual to ${\cal N} = 4$ super Yang-Mills on an anisotropic version of de Sitter space. Using the geodesic approximation for the two-point correlator of a high dimension operator, we found a signature of the bulk singularity. This correlator has a pole at the horizon scale when the points are separated in the direction corresponding to a negative Kasner exponent $p$. In our earlier paper \cite{Engelhardt:2014mea} we studied a particular example of this phenomenon for $p = -1/4$. We have shown here that the pole is always present when $p < 0$ and explained {\qf this is associated with geodesics probing the high curvature region} near the singularity. We have seen that the dual CFT state picked out by the geodesic approximation cannot be a standard state, but must be non-normalizable. 

How should we interpret this conclusion? Recall that a nontrivial bulk geometry corresponds to a state in the CFT with energy of order $N^2$. Different states for quantum fields on that bulk geometry correspond to adding excitations with energy of order one to this state. The CFT two-point function is the limit of a bulk two-point function in some state of the bulk quantum field. Presumably, there are well behaved bulk states which would lead to a CFT correlator in a normalizable state. So one conclusion is simply that 
the geodesic approximation does not select a reasonable state in the boundary theory and therefore can fail in the presence of cosmological singularities in AdS.\footnote{It is possible that the geodesic approximation also fails for some time dependent, nonsingular bulk geometries, although we do not know of any examples.}  If so, it remains an important open question to better understand the class of field theory states dual to the bulk singularity. In some sense, they should be  highly excited states containing many particles, because the bulk singularity corresponds to the asymptotic past on the de Sitter boundary.
 
An intriguing possibility is that there are significant finite $N$ corrections to the geodesic calculation of the correlator which smooths out the pole at the horizon. In other words, at finite $N$ the correlator in a normalizable state dual to the bulk geometry  might have large finite bumps at the horizon scale. One could then view the geodesic approximation as trying to reproduce a key feature of the exact answer. In the BCFT picture on the boundary this might correspond to specifying the state by introducing a fuzzy boundary. A scenario of this kind would mean however that the standard $1/N$ expansion does not have a continuous limit in the presence of our bulk singularity and would therefore be highly unusual.

It is natural to ask if there are other observables which could probe the singularity. One quantity that has attracted much recent attention is entanglement entropy. Following the seminal work \cite{Ryu:2006bv} by Ryu and Takayanagi, it was argued \cite{Hubeny:2007xt} that in a time dependent context, the entanglement entropy of a region $A$ in the dual field theory is given by the area of a bulk extremal surface which ends on the boundary of $A$. Unfortunately, it was shown in \cite{Wall:2012uf} that co-dimension two extremal surfaces cannot get close to the Kasner singularity.

If our bulk spacetime were  three-dimensional with metric \eqref{metric3d}, then the spacelike geodesics computed above would be the extremal surfaces needed for computing entanglement entropy. This raises an interesting puzzle. Given a region $A$ in the dual field theory, we can consider its domain of dependence $D[A]$. It has been shown \cite{Wall:2012uf} that the extremal surface has to stay outside the domain of influence of $D[A]$ in the bulk. Physically this is reasonable since  a local disturbance inside $A$ should not change the entanglement entropy and hence should not be able to change the area of the extremal surface.  However, when $p < 0$, we have seen that there are geodesics which approach the boundary and clearly lie inside the domain of influence. The resolution is that \eqref{metric3d} violates the null energy condition when $p < 0$ so the proof that the extremal surface lies outside the domain of influence does not apply. Interestingly, when $p>0$ the null energy condition is satisfied, and the spacelike geodesics which bend away from the singularity do stay outside the domain of influence.
 
Even though the entanglement entropy cannot directly probe the region of the bulk spacetime near the singularity, in light of our results it might still contain useful information about the CFT state. In particular, we mentioned earlier that the state selected by the geodesic approximation appears to contain pairs of correlated quasi-particles.  The presence of these quasi-particles should be manifest in the growth of the entanglement entropy in time. 
 
Another observable that could be investigated is the expectation value of Wilson loops. This is given by the area of two-dimensional extremal surfaces which are anchored on the loop on the boundary. It would be interesting to see if these expectation values have any unusual properties that can be associated with the bulk singularity.
 
%Of course the main open question is to better understand the field theory state dual to the bulk singularity. In particular,  what happens at finite $N$ to the signal of the bulk singularity that we have found, and is the pole really absent at finite coupling?

\vskip 1cm

\noindent{\bf Acknowledgements:} It is a pleasure to thank 
Adam Bzowski, Dalit Engelhardt, Sebastian Fischetti, Tom Hartman, Stefan Hollands, Don Marolf, and Steve Shenker
 for helpful discussions. This work is supported in part by the US NSF Graduate Research Fellowship under Grant No. DGE-1144085 and by NSF Grant No. PHY12-05500. NE would like to thank the KU Leuven Physics Department for their hospitality. TH thanks the KITP and the Physics Department at UCSB for their hospitality. The work of TH is supported in part by the Belgian National Science Foundation (FWO) grant G.001.12 Odysseus and by the European Research Council grant no. ERC-2013-CoG 616732 HoloQosmos.

\end{document}